\documentclass[epj,floatfix,twocolumn,showpacs,superscriptaddress]{revtex4}
\usepackage{graphicx}
\def\bea{\begin{eqnarray}}
\def\eea{\end{eqnarray}}
\def\be{\begin{equation}}
\def\ee{\end{equation}}

\begin{document}

\author{Didier~Poilblanc}
\affiliation{
  Laboratoire de Physique Th\'eorique, CNRS \& Universit\'e de Toulouse,
  F-31062 Toulouse, France }
\affiliation{
Institute of Theoretical Physics,
Ecole Polytechnique F\'ed\'erale de Lausanne,
BSP 720,
CH-1015 Lausanne,
Switzerland}

\author{Fabien~Alet}
\affiliation{
  Laboratoire de Physique Th\'eorique, CNRS \& Universit\'e de Toulouse,
  F-31062 Toulouse, France }

\author{Federico~Becca}
\affiliation{
  Laboratoire de Physique Th\'eorique, CNRS \& Universit\'e de Toulouse,
  F-31062 Toulouse, France }
\affiliation{CNR-INFM Democritos National Simulation Centre and
International School for Advanced Studies (SISSA),
I-34014 Trieste, Italy}

\author{Arnaud~Ralko}
\affiliation{
Institute of Theoretical Physics,
Ecole Polytechnique F\'ed\'erale de Lausanne,
BSP 720,
CH-1015 Lausanne,
Switzerland}

\author{Fabien~Trousselet}
\affiliation{
  Laboratoire de Physique Th\'eorique, CNRS \& Universit\'e de Toulouse,
  F-31062 Toulouse, France }

\author{Fr\'ed\'eric~Mila}
\affiliation{
Institute of Theoretical Physics,
Ecole Polytechnique F\'ed\'erale de Lausanne,
BSP 720,
CH-1015 Lausanne,
Switzerland}

\date{\today}
\title{Doping quantum dimer models on the square lattice}

\pacs{75.10.-b, 75.10.Jm, 75.40.Mg}
\begin{abstract}
A family of models is proposed to describe the motion of holes
in a fluctuating quantum dimer background on the square
lattice. Following Castelnovo {\it et al.} 
[Ann. Phys. (NY) {\bf 318}, 316 (2005)], a generalized
Rokhsar-Kivelson Hamiltonian at {\it finite doping} which can be
mapped on a {\it doped} interacting classical dimer model is constructed. 
A simple physical extension of this model is also considered. 
Using numerical computations and 
simple considerations based on the above exact mapping, we
determine the  phase diagram of the model
showing a number of quantum phases typical of a doped Mott insulator.
The two-hole
correlation function generically exhibits short-range or 
long-range algebraic correlations in the solid (columnar) 
and liquid (critical) phases of the model, respectively. 
Evidence for an extended region of a doped VBS phase
exhibiting holon pairing but {\it no} phase separation is given.
In contrast, we show that hole deconfinement occurs in 
the staggered dimer phase.

\end{abstract}
\maketitle

Soon after the discovery of cuprate superconductors with high
critical temperatures, Anderson suggested that the
Resonating Valence Bond (RVB) state is the relevant insulating
parent state that becomes superconducting under (arbitrary small)
hole doping~\cite{RVB}. Such a state can alternatively be viewed
as a spin liquid (SL), as it has no magnetic order and it does not
break any lattice symmetry. Since then, the search for exotic
SL in microscopic or effective models has been very active.

In quantum spin models, where magnetic
frustration suppresses long-range magnetic order, spin liquids 
often compete with quantum disordered states named ``Valence bond
solids'' (VBS) which break translation symmetry~\cite{VBS}. This is 
e.g. the case in the frustrated Heisenberg model with extended-range
antiferromagnetic (AF) interactions~\cite{reviews_frustrated}. In a VBS,
nearest-neighbor spins pair up in bond singlets which order e.g. along
columns or in a staggered arrangement. Hole doping 
has also been extensively studied 
in Mott insulators~\cite{reviews_HiTc}
and AF fluctuations have been identified as 
the glue for pairing. Unconventional pairing
upon doping  models exhibiting a VBS ground state has 
also been found~\cite{pairing_VBS}.

\begin{figure}
\centerline{\includegraphics*[angle=0,width=0.9\linewidth]{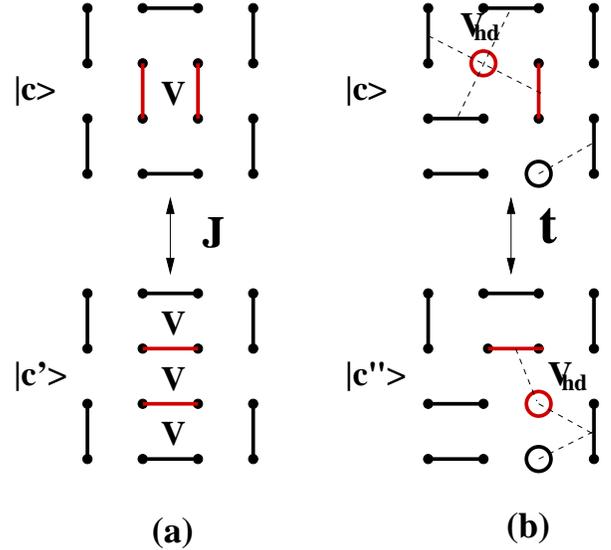}}
  \caption{\label{fig:configs}
(Color online) Pictures of the two quantum processes considered in this
paper. (a) Dimer flip $J$ within a plaquette. A dimer repulsion $V$ is
defined for all flippable plaquettes. (b) Hole hopping along a
plaquette diagonal. In this process a dimer ``rotates'' from a
vertical (horizontal) to a horizontal (vertical) bond. The
hole-dimer repulsion $V_{\rm hd}$ is defined on all the dashed lines.}
\end{figure}

In a pioneering work Rokhsar and Kivelson introduced a quantum
dimer model (QDM), a Hamiltonian acting in the space of
two-dimensional fully packed dimer configurations~\cite{RK} 
(generically called $|c\big>$). 
The dimer interaction $V$ and the dimer-flip process $J$ are
schematically depicted in Fig.~\ref{fig:configs}(a). 
As discussed in Ref.~\onlinecite{RK}, the QDM can be considered 
as the simplest 
effective model to describe quantum disordered phases similar to the
pseudo-gap phase
of the cuprate superconductors. In that respect, one might think of 
the dimer flip term as originated directly from super-exchange 
between copper spins. 
At the special point $V/J=1$, named
Rokhsar-Kivelson (RK) point, the ground state (GS) is exactly known and 
can be mapped onto the partition function of a classical 
dimer model~\cite{RK}. On the square lattice, the
dimer-dimer correlations are algebraic, decaying as $1/r^2$. This
``algebraic SL'' at the RK point is believed to be rather singular 
on the $V/J$ axis since, as shown e.g. by numerical 
calculations~\cite{QDM_num}, the GS is a VBS on both sides of it, a staggered 
phase for $V/J>1$, a columnar phase at attractive $V$ (i.e. $V<0$), separated
from the RK point by a small region of plaquette phase.
The case of non-bipartite lattices, where the RK point
has a gapped GS that shows fractional excitations~\cite{QDM_bis}, is
also of great interest.
Doping was introduced in Ref.~\onlinecite{RK} and studied further later
by Sylju{\aa }sen~\cite{doped_QDM} who computed dimer correlations and
the energy of two {\it static} monomers in a background of dimers and
dynamic holes. However, hole correlations of the dynamic holes 
themselves, have not been investigated so far.

In this article we construct 
various models of doped quantum dimers with the aim to study these correlations
as accurately as possible. 
Like the more ``microscopic'' t-J model, the lightly doped 
QDM also provides a realistic description of relevant
quantum disordered phases (which, in fact,
would be stabilized only at {\it finite} doping in the t-J model), 
while being much easier to handle numerically~\cite{note_spinons}. 
A generalized
RK Hamiltonian is introduced at {\it finite doping}.
This model can be mapped onto a {\it doped} interacting classical dimer model
enabling an efficient use of classical Monte
Carlo (MC)~\cite{Sandvik}. In addition, it offers a controllable parameter (the
effective temperature) to smoothly tune the system from a VBS to a 
liquid phase (even at zero doping). 
We also extend this
model to an enlarged physical space away from this so-called 
RK axis, where such a mapping is no longer valid~\cite{note} 
and where full quantum
computations such as Lanczos Exact Diagonalisation (ED)
and Green's function Monte Carlo (GFMC) 
are required. We provide evidence for hole
deconfinement~\cite{deconfinement} in the algebraic dimer phase. 
However, we also argue that phase separation occurs for low hole 
kinetic energy and, lastly, provide a complete phase diagram of the model. 

\begin{figure}
\centerline{\includegraphics*[angle=-90,width=0.9\linewidth]{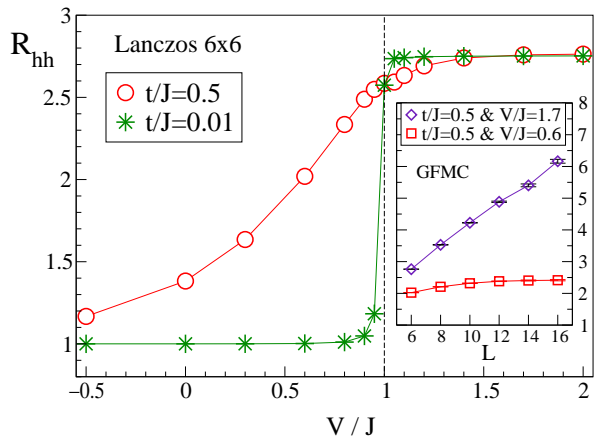}}
  \caption{\label{fig:hh_tJV}
(Color online)
Mean-squared separation $R_{\rm hh}$ between two holes in the t-J-V model 
as a function of the ratio $V/J$. Calculations are done on a 
periodic $6\times 6$ cluster for two values of the hopping $t/J$. 
Inset: Size dependence of $R_{hh}$ obtained from GFMC on $L\times L$ 
clusters for $t/J=0.5$ and two typical values of $V/J$
showing hole-hole deconfinement and confinement, respectively.}
\end{figure}

Let us first assume that holes are introduced by pairs on some of the dimer
bonds. Next, the simplest way to account for their motion~\cite{RK,doped_QDM} 
is to consider processes like 
the one depicted in Fig.~\ref{fig:configs}(b)
where a hole hops along a plaquette diagonal with some amplitude $t$. 
Note that here holes are really 
thought of as new charge degrees of freedom originating
e.g. from doping a Mott insulator.
A general form of doped QDM which operates in
a Hilbert space with a fixed hole number
can then be written as
\begin{equation}
H = \sum_c \epsilon_c |c\big>\big< c|  - J \sum_{(c,c')} |c'\big>\big< c| 
- t \sum_{(c,c'')} |c''\big>\big< c| \, ,
\label{eq:H}
\end{equation}
where the sum over $|c\big>$ includes all configurations with arbitrary
hole positions. The sum over $|c\big>$ and $|c'\big>$ 
extends on all pairs of doped dimer coverings (only)
differing by a single plaquette-flip [shown on
Fig.~\ref{fig:configs}(a)]. Similarly, the sum over $|c\big>$ 
and $|c''\big>$ extends to all pairs of doped dimer coverings
differing (only) by a single hole hopping along a diagonal of a
plaquette and a single dimer ``hop'' from a vertical (horizontal)
bond to a horizontal (vertical) bond. 
Such a process between two configurations
$|c\big>$ and $|c''\big>$ is depicted in Fig.~\ref{fig:configs}(b). 
As shown below, the role of the diagonal energies $\epsilon_c$ is crucial
and various choices will be discussed.
Whether mobile holes remain confined~\cite{confinement} 
or not (or whether they form bound states)
is the central issue of this study. 

Following Ref.~\onlinecite{RK},
we first start with the simple dimer interaction introduced above,
namely $\epsilon_c = \epsilon_c^0=V N_c$ where $N_c$ corresponds to the
number of flippable plaquettes in configuration $|c\big>$.
The properties of two 
holes~\cite{note_statistics} in
such a t-J-V model~\cite{doped_QDM} are studied here by ED 
and GFMC~\cite{GFMC} on finite clusters 
and the mean-squared
hole-hole distance $R_{\rm hh}=\sqrt{\big<r^2\big>}$ is shown in 
Fig.~\ref{fig:hh_tJV} as a function
of the dimer repulsion $V/J$ for fixed ratios $t/J$. A very abrupt
variation is observed at $V/J=1$ (especially at small $t$) showing
a clear confinement~\cite{confinement} at $V/J<1$, as expected in a VBS, 
and a large
value of the hole-hole separation for $V/J>1$, which scales linearly 
with system size (see inset of Fig~\ref{fig:hh_tJV}). In this case, the
undoped system is also a VBS, but a pure staggered dimer state
with no quantum fluctuations. It is therefore easy to see that
the two holes can freely move away from each other in opposite directions
along the same diagonal creating a string
of dimers at 90$^\circ$ from the background at no energy cost~\cite{Shannon}.
On the contrary, in a plaquette or columnar VBS the
energy cost grows linearly with the string length leading to 
confinement. Note that it was argued that confinement is lost for a
sufficiently high fraction of holes (monomers)~\cite{doped_QDM}.

\begin{figure}
\centerline{\includegraphics*[angle=0,width=0.9\linewidth]{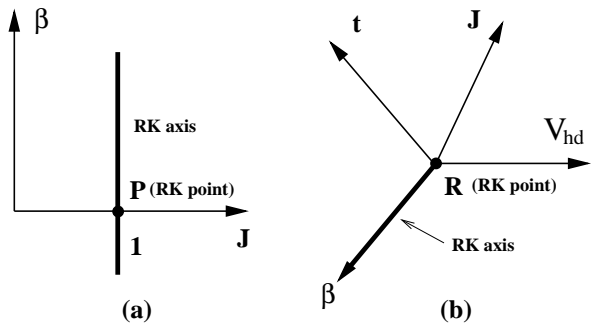}}
  \caption{\label{fig:models}
Schematic picture of the 
parameter space of the models considered here; (a) J-$\beta$ model (undoped) 
and (b) t-J-$\beta$ model (doped). The RK axes (thick lines parametrized 
by $\beta$) are defined by $J=1$ (undoped case) and by $J=t=V_{\rm hd}=1$
(doped case). The points P and R correspond to the origins at $\beta=0$ 
on these axes.}
\end{figure}

In order to construct more general doped QDM, we 
briefly re-examine the undoped case (i.e. zero doping for which
the $t$ term of Eq.~(\ref{eq:H}) is irrelevant)
and introduce a simple construction that extends the undoped RK
point to an infinite axis. Following Castelnovo {\it et al.}~\cite{mapping}, 
we define the J-$\beta$ Hamiltonian by introducing new
diagonal energies as,
\begin{equation}
\label{eq:J-beta_diag}
\epsilon_c  = \epsilon_c^{\rm flip} 
=V \sum_{c'(c)} \exp{\{-\frac{1}{2} \beta V_{\rm cl} (N_{c'} - N_c)\}}
\end{equation}
where the sum extends on the dimer covering $|c'\big>$
differing from $|c\big>$ by a single plaquette-flip [shown on
Fig.~\ref{fig:configs}(a)].
In the $\beta\rightarrow 0$ limit, the expression of $\epsilon_c^{\rm flip}$
reduces to $\epsilon_c^0$ and the significance of $V$ becomes clear.
Hereafter, $V=1$ sets the energy scale. 
$V_{\rm cl}$ corresponds to a {\it classical} dimer-dimer interaction. 
We restrict ourselves to the attractive case and use units 
for $\beta$ such that $V_{\rm cl}=-1$. A cartoon of the Hamiltonian manifold
parametrized by $(J,\beta)$ is shown in Fig.~\ref{fig:models}(a).
For $J=1$ it is easy to
check that the GS is simply given by $\frac{1}{\sqrt{Z}} \sum_c
\exp{(-\frac{1}{2}\beta V_{\rm cl} N_c)} |c\big>$ with energy $E_0=0$,
where the normalization factor $Z=\sum_c \exp{(-\beta V_{\rm cl} N_c)}$ can
be considered as a partition function of a 
classical interacting dimer model~\cite{class_DM}.
We have checked by ED data of a $8 \times 8$ 
cluster within its fully symmetric space-group 
irreducible representation, 
that the specific heat [defined as $\beta^2(\big<N_c^2\big>-\big<N_c\big>^2)$]
is very close to the MC results
obtained for a very large cluster (not shown).
The model displays a Kosterlitz-Thouless (KT) transition~\cite{KT}
at $\beta=\beta_{\rm KT}\simeq 1.536$ between a critical 
phase at $\beta<\beta_{\rm KT}$ (with $\beta$ varying exponents)
and a columnar dimer phase~\cite{class_DM}. 
Independently from our investigation, a similar mapping was derived and
the properties of this critical phase were investigated with transfer-matrix 
techniques~\cite{mapping2}. 

To investigate the expected
confinement-deconfinement transition~\cite{deconfinement} for holes at the
KT transition, let us now generalize the construction by Castelnovo 
{\it et al.}~\cite{mapping} to finite doping.
We define the t-J-$\beta$ Hamiltonian by adding a second diagonal term to
the one~(\ref{eq:J-beta_diag}) of the J-$\beta$ Hamiltonian,
$\epsilon_c=\epsilon_c^{\rm flip}+\epsilon_c^{\rm hop}$,
\begin{equation}
\label{eq:kinetic}
\epsilon_c^{\rm hop}=V_{\rm hd}\sum_{c''(c)} \exp{\{- \frac{1}{2} 
\beta V_{\rm cl} (N_{c''}- N_c)\}}
\end{equation}
where the sum now extends on the doped dimer coverings
$|c^{''}\big>$ connected to $|c\big>$ 
by a $t$-process [see again Fig.~\ref{fig:configs}(b)]. 
Note that this new term, as the $t$-term of (\ref{eq:H}), 
scales like the hole concentration.
$V_{\rm hd}$ is a new energy scale which naturally makes sense in the $\beta
\rightarrow 0$ limit discussed below. The term (\ref{eq:kinetic}) 
is of central importance as for
$t=V_{\rm hd}$ and $J=1$ the GS can again be written as
$\frac{1}{\sqrt{Z}}\sum_c \exp{(-\frac{1}{2}\beta V_{\rm cl} N_c)}
|c\big>$. 
In analogy with the undoped case, we can again define a 
RK axis which runs along an orthogonal
direction to the three-dimensional parameter space spanned 
by $J$, $t$, and $V_{\rm hd}$.
This RK axis is given by $J=1$, $t=V_{\rm hd}$ and parametrized by $\beta$. 
A cartoon picture of this set of Hamiltonians is shown in
Fig.~\ref{fig:models}(b).
Since the hopping term (for $t\ne 0$) couples all
topological symmetry sectors the GS with energy
$E_0=0$ becomes unique. 
Interestingly,
the procedure followed here can be generalized to more
complicated dimer or hole kinetic off-diagonal processes.

\begin{figure}
\centerline{\includegraphics*[angle=-90,width=0.9\linewidth]{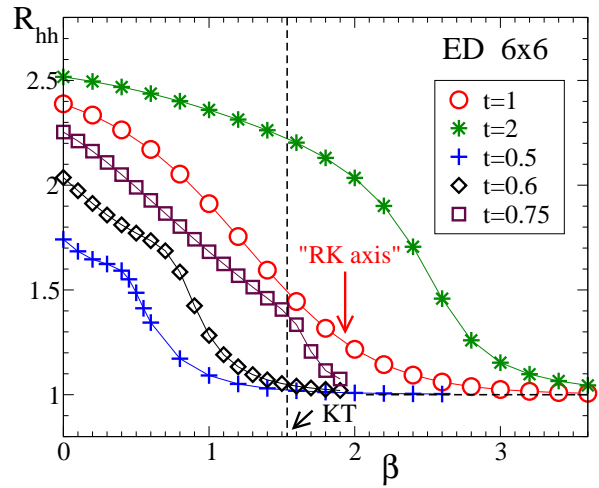}}
  \caption{\label{fig:hh_tJ-beta}
(Color online)
Mean-squared separation between two holes in the t-J-$\beta$ model 
as a function of $\beta$. ED data obtained on a $6\times 6$ 
periodic cluster are shown for several values of the hopping $t$ and
$J=1$. The dashed line denotes the location of the KT transition at
$\beta_{KT}\sim 1.536$. For $t=0.5$ and $t=0.6$ the ''kinks''
signal the appearance of a two-hole bound state.}
\end{figure}

The $\beta\rightarrow 0$ limit is of special interest. 
As seen above, the diagonal term (\ref{eq:J-beta_diag})
reduces to a dimer-dimer repulsion of magnitude $V$ (set to 1) and 
the {\it undoped} 
RK point is recovered for $J=1$ (when holes are not present). 
Similarly, the second diagonal contribution (\ref{eq:kinetic}) 
reduces to a dimer-hole interaction on a
plaquette [with definition
given pictorially in Fig.~\ref{fig:configs}(b)] of magnitude $V_{\rm hd}$. 
The $\beta=0$ limit therefore gives rise to a large class of physical 
Hamiltonians parametrized by arbitrary
magnitudes of $J$, $t$ and  $V_{\rm hd}$ (measured in units of $V=1$). 
A complete investigation of this model is left for a future 
study~\cite{complete} and, in the following, we restrict
ourselves to $J=V$ ($=1$ for convenience) so that the ``distance'' 
from the RK-axis will be controlled by the deviation of $t$ from $V_{\rm hd}$. 

\begin{figure}
\centerline{\includegraphics*[angle=0,width=0.9\linewidth]{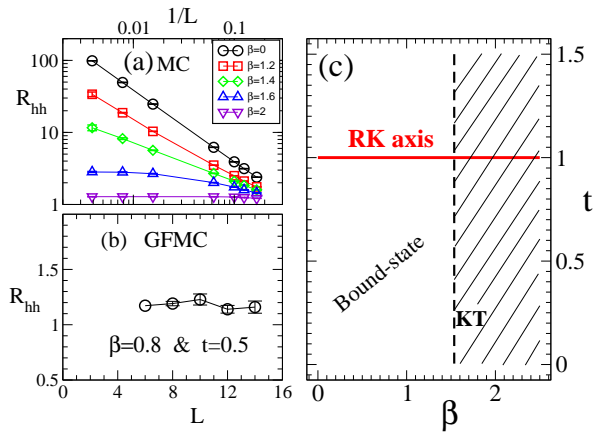}}
  \caption{\label{fig:phase_diag}
(Color online) Results for 2 holes in 
the  t-J-$\beta$ model ($J=V$($=1$)) for  $V_{\rm hd}=V$($=1$). 
Size dependence of
$R_{hh}$ by (a) classical MC at $t=1$ for various values of
$\beta$ (log-log scale), (b) GFMC at $\beta=0.8$ 
and $t=0.5$ (linear scale). 
(c) Conjectured phase diagram for two holes in
the thermodynamic limit for $J=V=V_{\rm hd}$($=1$) vs $\beta$ and $t$. 
The dashed region
corresponds to the confined phase for $\beta>\beta_{\rm KT}$.
}
\end{figure}

We first start with the case of two holes in the t-J-$\beta$
model.
The hole-hole correlations have been computed by ED of a $6\times 6$ cluster 
for arbitrary $t$ and, for convenience, for $V_{\rm hd}=1$. 
Results are shown in Fig.~\ref{fig:hh_tJ-beta}.
The ED results show a rather smooth variation of $R_{\rm hh}$
across the transition at $\beta_{\rm KT}$.
However, the finite size scaling 
at $t=1$ obtained by classical MC in Fig.~\ref{fig:phase_diag}(a)
shows a clear qualitative change of behavior at the KT transition:
while $R_{hh}$ remains finite in the confined phase,
it diverges as a power law in the critical phase in agreement with 
Ref.~\onlinecite{class_DM}. 
For $t=0.5$ and $t=0.6$ we observe a kink in the ED data of
Fig.~\ref{fig:hh_tJ-beta}. Moreover, for $t\le 0.25$ $R_{\rm hh}$
remains always very close to 1, even when $\beta\rightarrow 0$ (not shown).
This signals the
appearance of a two-hole bound state within the critical SL 
phase for $\beta$ below $\beta_{\rm KT}$.
This scenario is supported by the GFMC data on larger systems shown in 
Fig.~\ref{fig:phase_diag}(b).
We have checked by GFMC that this bound state, in fact, 
persists up to $t=1$ (=$V_{hd}$).
These results are
summarized in a phase diagram for two holes in Fig.~\ref{fig:phase_diag}(c).

\begin{figure}
\centerline{\includegraphics*[angle=0,width=0.9\linewidth]{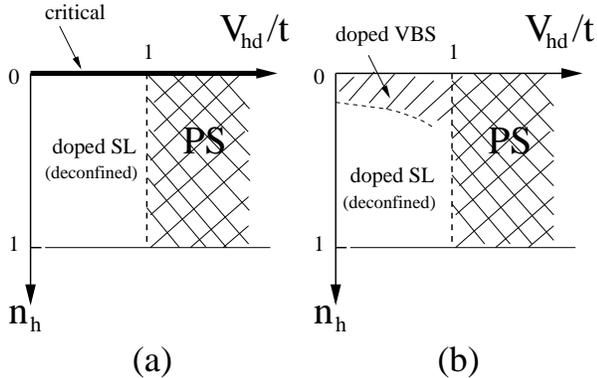}}
  \caption{\label{fig:phase_diag3}
Phase diagrams of the t-J-$\beta$ for $V=J$ 
vs hole density and $V_{hd}/t$ ratio .
''PS'' stands for ''phase separation''.
(a) Phase diagram for $\beta<\beta_{\rm KT}$;
(b) Phase diagram for $\beta>\beta_{\rm KT}$. The boundary between the
confined (doped VBS) and deconfined regions is schematic. 
}
\end{figure}

Lastly, using simple arguments, we construst the phase diagram of the
doped QDM for $V=J$ (and bosonic holes). 
The previous finding that two holes pair up in 
some region of parameter space implies that either (i) a
Q=2e superconducting state or (ii) a phase separated state 
occurs at finite doping. For $V_{ht}=t$, the inverse 
compressibility $\kappa^{-1}$ (proportional to the second derivative of the
energy per site w.r.t the hole density) identically vanishes 
since the GS energy vanishes on the RK axis for all number of holes. 
In addition, since $V_{\rm hd}$
acts effectively as an attraction between holes (or dimers), $\kappa^{-1}$
should be a monotonous function of the ratio $V_{\rm hd}/t$
as checked numerically~\cite{complete}. These simple considerations
imply that $\kappa^{-1}<0$ ($>0$) for $t<V_{\rm hd}$ ($t>V_{\rm hd}$)
and, then, the phase separation boundary is exactly 
given by $t=V_{\rm hd}$ for all 
hole density and $\beta$. Fig.~\ref{fig:phase_diag3} shows the phase diagram
of the model for $J=V$ as a function of $t$,  $V_{\rm hd}$ and hole density. 
For $\beta<\beta_{\rm KT}$, holons are deconfined in the 
SL phase. We also expect a similar phase diagram
for $\beta>\beta_{\rm KT}$ (Fig.~\ref{fig:phase_diag3}(b)) 
although the critical line (at $n_h=0$)
should be replaced by an extended (hole pair) 
region with VBS columnar order which survives
up to a critical doping. However, the boundary of the 
phase separated region remains unchanged.
The exact curve for the VBS/SF boundary remains
to be investigated in more details.

In conclusion, we have introduced a class of simple doped QDM on the
square lattice which, we believe, provide insights
on the physical quantum disordered phases of real materials.
An exact mapping onto a doped classical dimer model 
(characterized by an inverse temperature $\beta$) 
can be realized along a one-dimensional manifold of this 
multi-dimensional space. Using numerical computations and 
simple considerations based on the above exact mapping, we
determine the complete phase diagram of the model (at $V=J$)
showing a number of interesting physical phenomena that, we believe,
could be generic in the vicinity of a Mott insulator, beyond the 
framework of QDM. It is found that two doped holons 
are confined in the columnar VBS phase.
At finite doping, evidence for an extended region of a metallic VBS phase
exhibiting holon pairing {\it without} 
phase separation is given.
In contrast, we find that the algebraic dimer-dimer correlations of the
(undoped) critical (quasi-ordered) phase do not provide the
confinement of two injected holons.
Furthermore, as shown in the case of the staggered dimer phase,
we point out that holon deconfinement can even occur in
a VBS.

We thank IDRIS (Orsay, France) for allocation of CPU-time on the
NEC-SX5 supercomputer. We are indebted to P.~Pujol for 
many useful suggestions and thank G.~Misguich for
interesting discussions. D.P. also thanks S.~Kivelson for useful
comments. This
work was partially supported by the Swiss National Fund and by
MaNEP. F.B. is partially supported by COFIN 2004, COFIN 2005 (Italy),
and by CNRS (France).


\begin{thebibliography}{}
\bibitem{note_spinons} Fermionic spin-1/2 spinon
degrees of freedom are not included in the QDM, which is justified
in pseudo- or spin gap phases for energies below
the gap.
\bibitem{confinement} By ``confinement'' we implicitly mean
(based on qualitative arguments) that the onset of the two hole 
continuum is pushed up to (strictly speaking) infinite energy. 
This is fundamentally different from a simple two hole bound 
state which can appear in a liquid dimer state.
\end{thebibliography}
\end{document}